\documentclass[twocolumn,showpacs,preprintnumbers,amsmath,amssymb,prl,aps,superscriptaddress]{revtex4-1}
\usepackage{graphicx}
\usepackage{amsmath}
\usepackage{verbatim}

\newcommand{\kHz}{\mathrm{kHz}}
\newcommand{\MHz}{\mathrm{MHz}}
\newcommand{\GHz}{\mathrm{GHz}}

\newcommand{\ns}{\mathrm{ns}}

\newcommand{\Qone}{Q_{1}}

\newcommand{\Qtwo}{Q_{2}}

\newcommand{\Q}{Q}
\newcommand{\A}{A}
\newcommand{\B}{B}

\newcommand{\Hone}{H_{1}}
\newcommand{\Htwo}{H_{2}}
\newcommand{\Bus}{B}

\newcommand{\Ma}{M_A}
\newcommand{\Manorm}{\tilde{M}_A}
\newcommand{\Mq}{M_Q}

\newcommand{\tswap}{\tau_{\mathrm{s}}}
\newcommand{\twait}{\tau_{\mathrm{w}}}

\newcommand{\Xa}{X_A}
\newcommand{\Ya}{Y_A}
\newcommand{\Za}{Z_A}

\newcommand{\Xq}{X_Q}
\newcommand{\Yq}{Y_Q}
\newcommand{\Zq}{Z_Q}

\newcommand{\xq}{x_Q}

\newcommand{\xa}{x_A}
\newcommand{\ya}{y_A}

\newcommand{\ket}[1]{\left\lvert #1 \right\rangle}
\newcommand{\bra}[1]{\left\langle #1 \right\rvert}
\newcommand{\avg}[1]{\left\langle #1 \right\rangle}
\newcommand{\avgb}[1]{\langle #1 \rangle}

\newcommand{\ketq}[1]{\left\lvert #1_{Q}\right\rangle}

\newcommand{\braket}[2]{\langle #1| #2\rangle}

\newcommand{\ECOneTwo}{E_{C1(2)}}

\newcommand{\gH}{g_{H}}
\newcommand{\gB}{g_{B}}

\newcommand{\ChiA}{\chi_{\mathrm{A}}}

\newcommand{\W}{W}
\newcommand{\Wm}{W_{\mathrm{m}}}

\newcommand{\Bpm}{B_{\pm}}
\newcommand{\Bp}{B_{+}}
\newcommand{\Bm}{B_{-}}

\newcommand{\wmaxOneTwo}{\omega^{\mathrm{max}}_{Q1(2)}}

\newcommand{\wHOneTwo}{\omega_{H1(2)}}
\newcommand{\kHOneTwo}{\kappa_{H}}
\newcommand{\wbus}{\omega_{B}}
\newcommand{\kbus}{\kappa_{B}}

\newcommand{\Rmnum}[1]{\expandafter\@slowromancap\romannumeral #1@}

\begin{document}
\title{Partial-measurement back-action and non-classical weak values in a superconducting circuit}
\author{J.~P.~Groen}
\author{D.~Rist\`e}
\affiliation{Kavli Institute of Nanoscience, Delft University of Technology, P.O. Box 5046,
2600 GA Delft, The Netherlands}
\author{L.~Tornberg}
\affiliation{Microtechnology and Nanoscience, MC2, Chalmers University of Technology, SE-412 96 Goteborg, Sweden}
\author{J.~Cramer}
\affiliation{Kavli Institute of Nanoscience, Delft University of Technology, P.O. Box 5046,
2600 GA Delft, The Netherlands}
\author{P.~C.~de Groot}
\affiliation{Kavli Institute of Nanoscience, Delft University of Technology, P.O. Box 5046,
2600 GA Delft, The Netherlands}
\affiliation{Max Planck Institute for Quantum Optics, Garching 85748, Munich, Germany}
\author{T.~Picot}
\affiliation{Kavli Institute of Nanoscience, Delft University of Technology, P.O. Box 5046,
2600 GA Delft, The Netherlands}
\affiliation{Laboratory of Solid-State Physics and Magnetism, KU Leuven, Celestijnenlaan 200D, 3001 Leuven, Belgium}
\author{G.~Johansson}
\affiliation{Microtechnology and Nanoscience, MC2, Chalmers University of Technology, SE-412 96 Goteborg, Sweden}
\author{L.~DiCarlo}
\affiliation{Kavli Institute of Nanoscience, Delft University of Technology, P.O. Box 5046,
2600 GA Delft, The Netherlands}
\date{\today}

\begin{abstract}
We realize indirect partial measurement of a transmon qubit in circuit quantum electrodynamics by interaction with an ancilla qubit and projective ancilla measurement with a dedicated readout resonator. Accurate control of the interaction and ancilla measurement basis allows tailoring the measurement strength and operator.  The tradeoff between measurement strength and qubit back-action is characterized through the distortion of a qubit Rabi oscillation imposed by ancilla measurement in different bases. Combining partial and projective qubit measurements, we provide the solid-state demonstration of the correspondence between a non-classical weak value and the violation of a Leggett-Garg inequality.
\end{abstract}

\pacs{03.67.Lx, 42.50.Dv, 42.50.Pq, 85.25.-j}
\maketitle

Quantum measurement involves a fundamental tradeoff between information gain and disturbance of the measured system that is traceable to uncertainty relations~\cite{Braginsky95}. The back-action, or kick-back, is a non-unitary process that depends on the measurement result and pre-measurement system state. Thought experiments in the 1980's unveiled paradoxes where the back-action of multiple measurements of one system puts quantum mechanics at odds with macrorealism (MAR)~\cite{Leggett85}, a set of postulates distilling our common assumptions about the macroscopic world. The paradoxes include the violation of  Bell inequalities in time, also known as Leggett-Garg inequalities (LGIs)~\cite{Leggett85}, the non-classicality of weak values~\cite{Aharonov88}, and the three-box problem~\cite{Aharonov91}.

Steady developments in control of single systems for quantum information processing have opened the road to fundamental investigations of back-action with photons~\cite{Resch04,Pryde05,Goggin11}, superconducting circuits~\cite{PalaciosLaloy09}, and semiconductor spins~\cite{Waldherr11, Knee12, George12}.  Experiments decidedly favor quantum mechanics over MAR, although loopholes exist in each case. Moving beyond fundamental investigation, the emergent field of quantum feedback control~\cite{Wiseman09} balances the tradeoff between information gain and back-action, finding applications in quantum error correction~\cite{Branczyk07}, qubit stabilization~\cite{Wang01,Vijay12}, and state discrimination~\cite{Waldherr12}, for example, where partial measurements are preferred over maximally-disturbing projective ones.

In superconducting circuits, variable-strength measurement was first demonstrated for a Josephson phase qubit~\cite{Katz06}. Although destructive for the qubit for one of two measurement outcomes, the method allowed probabilistic wavefunction uncollapse~\cite{Korotkov06} by two sequential partial measurements, firmly demonstrating that back-action is phase-coherent~\cite{Katz08}. Very recently~\cite{Hatridge13}, partial measurement of a transmon qubit was realized in circuit quantum electrodynamics (cQED)~\cite{Wallraff04,Blais04} by probing  transmission through a dispersively-coupled cavity. In this case the measurement strength was controlled through the number of photons in the probe signal.

In this Letter, we demonstrate a non-destructive, variable-strength measurement of a transmon qubit that is  based on controlled interaction with an ancilla qubit and projective ancilla measurement [Fig.~1(b)]. The key advantage of ancilla-based indirect measurement is the possibility to accurately tailor the measurement by control of the interaction step and choice of ancilla measurement basis.  The kick-back of variable-strength measurements on the qubit is investigated by conditioning tomographic qubit measurements on the result of ancilla measurements in different bases, showing close agreement with theory. By combining partial and projective measurements, we observe non-classical weak values, LGI violations, and show their predicted correspondence~\cite{Williams08,*Williams09,Goggin11} in a solid-state system. The ancilla-based indirect measurement here demonstrated will be extendable to realization of four-qubit parity measurements needed for modern error correction~\cite{Fowler12}.

\begin{figure}
\includegraphics[width=\columnwidth]{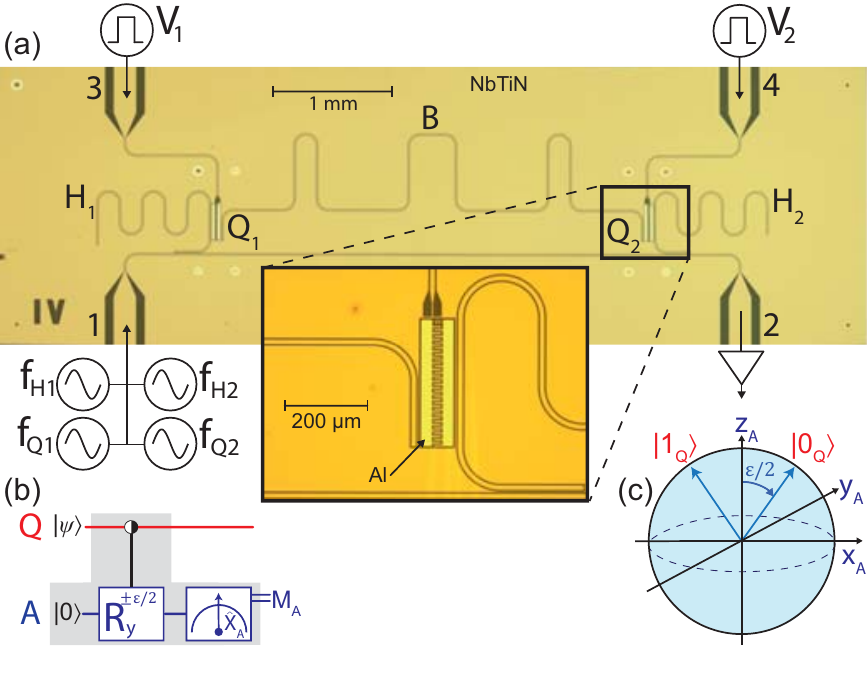}
\caption{(color online). (a) Two-transmon, three-resonator cQED processor. Quarter-wave resonators $\Hone$ and $\Htwo$ allow individual readout of qubits $\Qone$ and $\Qtwo$ via a common feedline. A resonator bus $\Bus$ (single-photon quality factor $210,000$) couples to both qubits. Local flux-bias lines (ports 3 and 4) allow independent tuning of qubit transition frequencies with $\sim\!1~\GHz$ bandwidth~\cite{DiCarlo09}. (b) Scheme for two-step indirect measurement of one qubit ($\Q$) through partial entanglement with an ancilla qubit ($\A$) followed by projective measurement of $\A$. (c) Bloch-sphere illustration of the evolution of $\A$ during the interaction step, for $\Q$ in $\ketq{0}$ and $\ketq{1}$.}
\end{figure}

Our planar cQED device~\cite{SOM} consists of two transmon qubits ($\Qone$ and $\Qtwo$) coupling jointly to a bus resonator ($\Bus$) and separately to dedicated resonators ($\Hone$ and $\Htwo$) used for projective readout [Fig.~1(a)]. Local flux-bias lines allow control of individual qubit transition frequencies on nanosecond timescales. All microwave pulses for individual qubit control and readout are applied through a common feedline coupling to $\Hone$ and $\Htwo$. Qubit $\Qone~(\Qtwo)$ has charging energy $\ECOneTwo = 300~(330)~\MHz$ and maximum transition frequency $\wmaxOneTwo/2\pi=7.37~(6.55)~\GHz$.  The bus $\Bus$ has fundamental frequency $\wbus/2\pi = 6.15~\GHz$ and intrinsic linewidth $\kbus/2\pi = 30~\kHz$, and couples to both qubits with $\gB/2\pi = 36~\MHz$. Resonant swaps between either qubit and $\Bus$ are realized in $\tswap=7~\ns$. The readout resonators have fundamental frequencies $\wHOneTwo/2\pi = 7.13~(7.24)~\GHz$ and coupling-limited linewidth $\kHOneTwo/2\pi=9~\MHz$, and couple to their respective qubit with $\gH/2\pi = 92~\MHz$. Projective dispersive readout of $\Qone$ ($\Qtwo$) is performed by pulsed homodyne detection of feedline transmission at the resonance frequency of $\Hone~(\Htwo)$ with $\Qone~(\Qtwo)$ in the ground state. Digitization at the optimal threshold gives $85\%$ ($94\%$) single-shot readout fidelity, respectively.

\begin{figure}
\includegraphics[width=\columnwidth]{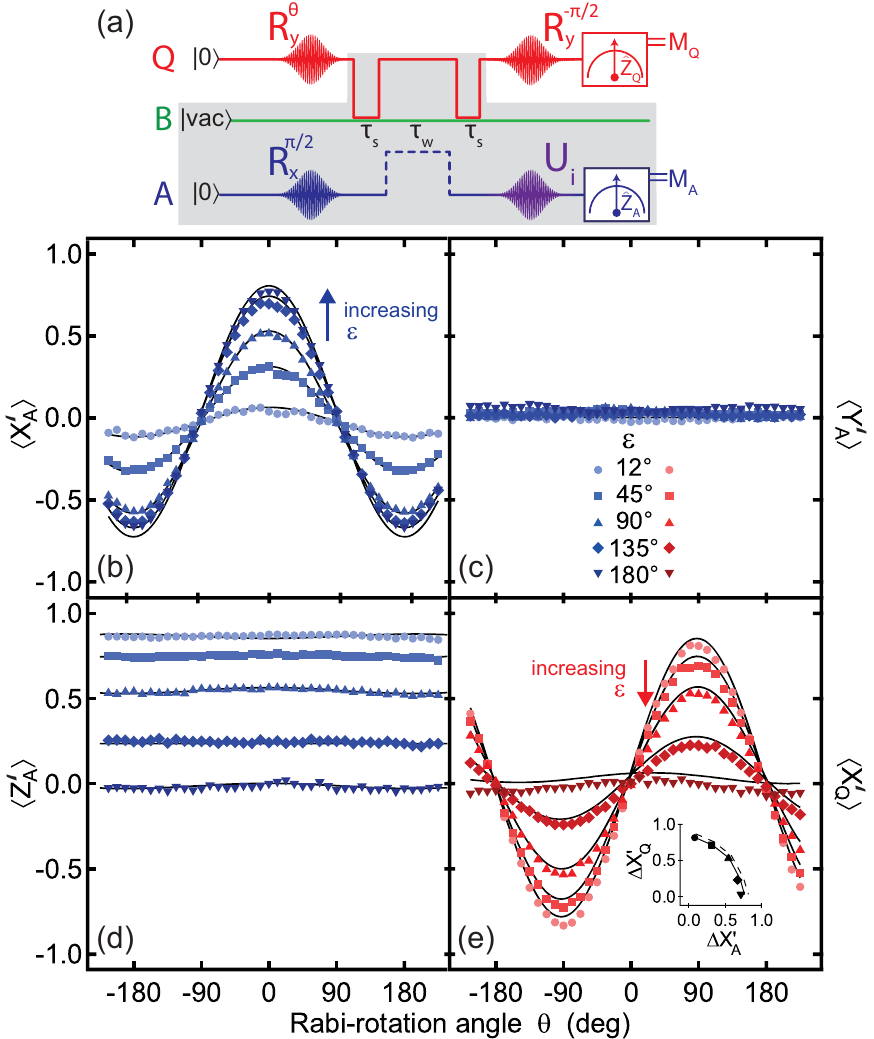}
\caption{(color online). (a) Pulse sequence realizing and testing the indirect measurement [Fig.~1(b)] of qubit $\Q=\Qone$ with ancilla $\A=\Qtwo$. $\Q$ is first prepared in state  $\ket{\theta}=\cos(\theta/2) \ketq{0}+ \sin(\theta/2) \ketq{1}$. (b-d) Ensemble-averaged ancilla measurement in the (b) $\Xa$, (c) $\Ya$, and (d) $\Za$ bases, achieved using pre-measurement rotation $U_i=R_{\ya}(-\pi/2)$, identity, and $R_{\xa}(-\pi/2)$, respectively. (e) Partial-measurement back-action reduces the contrast in the final projective measurement of $\Q$. Ideally, $\avg{\Xq'}=\avg{\Zq}\cos(\epsilon/2)$, independent of $\Ma$ basis ($\Xa$ here, see~\cite{SOM} for the three bases). Inset: Parametric plot of oscillation amplitudes $\Delta \Xq'$ and $\Delta \Xa'$.  $\Delta \Xq'$~$(\Delta \Xa')$ decreases (increases) as $\epsilon$ increases in the range $\left[0^\circ,180^\circ\right]$. Solid (dashed) curves correspond to the model with (without) decoherence during the gate.}
\end{figure}

The interaction step in the indirect measurement is a $y$ rotation of the ancilla ($\A=\Qtwo$) by $\pm \epsilon/2$, with positive (negative) sign for $\Q=\Qone$ in $\ketq{0}$~$(\ketq{1})$ [Fig.~1(c)]. The angle $\epsilon$ sets the measurement strength. Note that $\epsilon=180^\circ$ makes the measurement projective, as in this case $\A$ evolves to orthogonal states for $\ketq{0}$ and $\ketq{1}$. The $\Q$-dependent $y$ rotation of $\A$ is achieved by dressing a controlled-$z$ rotation with pre- and post-rotations on $\A$ [Fig.~2(a)]. The controlled-$z$ rotation is a three-step process: a resonant swap transferring the state of $\Q$ to $\B$, a photon-controlled $z$ rotation of $\A$, and a resonant swap from $\B$ back to $\Q$. The acquired two-qubit phase $2|\ChiA|\twait=\epsilon$ is calibrated by varying the wait time $\twait$ and the Stark shift on $\A$ induced by one photon in $B$~\cite{SOM}. Single-qubit phases are nulled by phase-shifting all qubit microwave drives after flux pulsing, realizing virtual $z$-gates~\cite{Steffen00, Reed12}.

We characterize the interaction step by performing tomographic measurements of $\Q$ and $\A$ after the interaction, with $\Q$ nominally prepared in the superposition state $\ketq{\theta} = \cos(\theta/2) \ketq{0}+ \sin(\theta/2) \ketq{1}$ and $\A$ and $\B$ in the ground state. As in this part we focus purely on the interaction, we correct for readout errors using standard calibration procedures~\cite{Wallraff05, SOM}. Ideally, $\avg{\Xa'}=\avg{\Zq} \sin(\epsilon/2)$, $\avg{\Ya'}=0$ and $\avg{\Za'}= \cos(\epsilon/2)$ [(un-)primed notation denotes the (pre-) post-interaction state]. These dependencies are well reproduced in the data for all choices of $\epsilon$ [Figs.~2(b-d)]. Measuring in either the $\Ya$ or $\Za$ basis yields no information about the initial state of $\Q$, as expected. We also measure the post-interaction contrast $\Delta \Xq'$ of the $\Q$ Rabi oscillation [Fig.~2(e)] and compare it to the contrast $\Delta \Xa'$ of the interaction-induced oscillation in $\A$. As $\epsilon$ increases, $\Delta \Xq'$  decreases while $\Delta \Xa'$ increases. Ideally, the quadrature sum $\sqrt{(\Delta\Xq')^2+(\Delta\Xa')^2}=1$ for any $\epsilon$. We observe a monotonic decrease from $0.82$ at $\epsilon=12^\circ$ to $0.72$ at $\epsilon=180^\circ$ [Fig.~2(e) inset]. This decrease is reproduced by a master equation simulation that includes $4\%$ residual excitation  and  measured decoherence rates for each element~\cite{SOM}.

\begin{figure}
\includegraphics[width=\columnwidth]{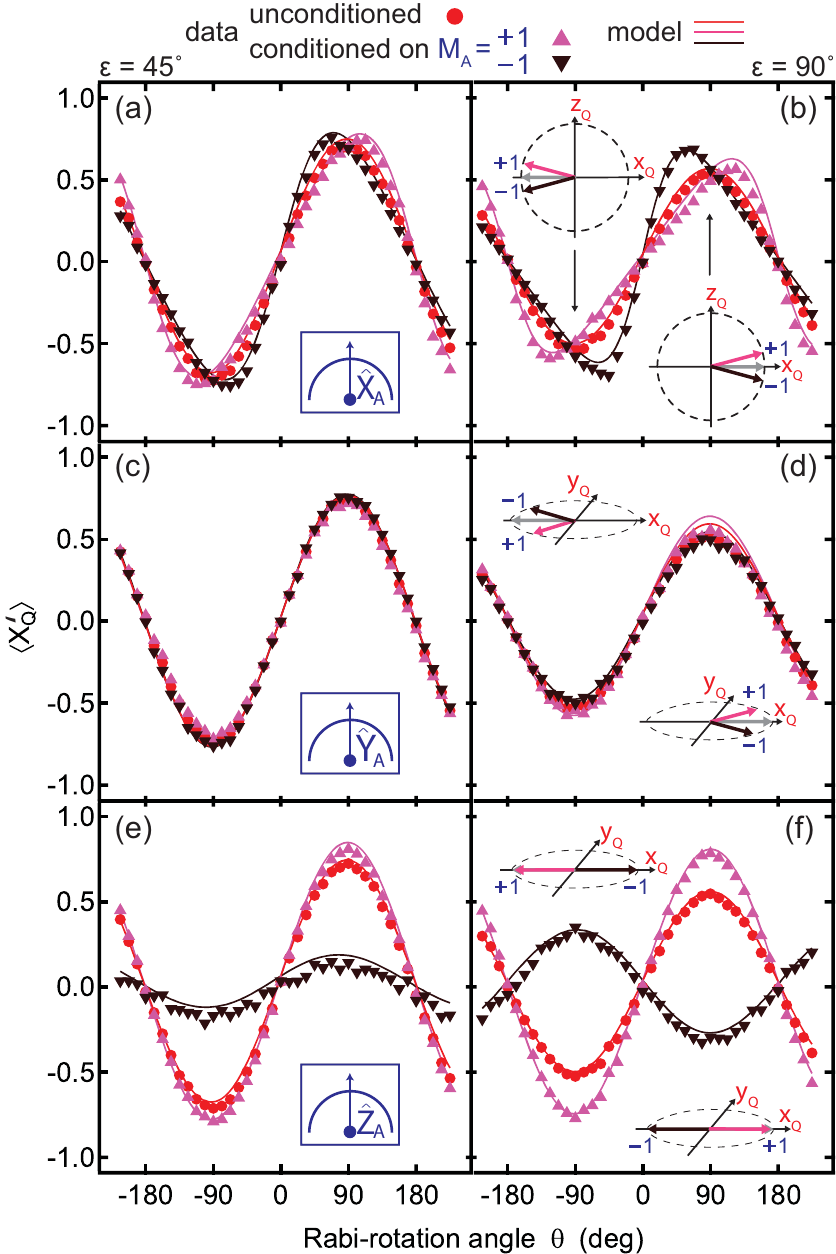}
\caption{(color online). Partial-measurement kick-back.
The kick-back on qubit $\Q$ induced by partial measurement depends on the interaction strength $\epsilon$, the ancilla measurement basis, and measurement result $\Ma$. Left and right panels correspond to $\epsilon=45^\circ$ and $90^\circ$, respectively. (a,b) Conditioning on the result $\Ma$ of ancilla measurement in the $\Xa$ basis reveals distorted Rabi oscillations of $\Q$. A positive (negative) result retards (advances) the oscillation for $\theta\in \left[0^\circ,180^\circ\right]$ and advances (retards) it for $\theta\in\left[0^\circ, 360^\circ\right]$. (c,d) Distortions for measurement in the $\Ya$ basis. In this case, the kick-back is a $z$ rotation by  $\pm\epsilon/2$,  causing an identical reduction of contrast in the conditioned Rabi oscillations.
(e,f) Distortions for measurement in the $\Za$ basis. Ideally, $\Ma=+1$ has no kick-back on $\Q$, while $\Ma=-1$ causes a $z$ rotation of $\pi$. The lower contrast observed for $\Ma=-1$ is due to readout errors (see text and~\cite{SOM} for details).
}
\end{figure}

We now investigate the quantum kick-back of partial measurements by performing partial state tomography of $\Q$ conditioned on the result of the ancilla measurement $\Ma=\pm1$ in different bases. Results in Fig.~3 show the partial-measurement induced distortion of a Rabi oscillation of $\Q$ for $\epsilon = 45^\circ$ and $90^\circ$ (see SOM~\cite{SOM} for other $\epsilon$ values) and measurement of $\Xa$, $\Ya$, and $\Za$. For $\Xa$, $\Ma=\pm 1$ kicks $\Q$ toward the north (south) pole of the Bloch sphere. Ideally, the Bloch vector polar angle transforms as  $\theta \rightarrow \theta'$, with $\tan(\theta'/2)=\tan(\theta/2)[1-\Ma\tan(\epsilon/4)]/[1+\Ma\tan(\epsilon/4)]$, while the azimuthal angle is conserved. Readout errors decrease the amplitude of the conditioned curves. The difference in the amplitudes is due to asymmetric readout errors~\cite{SOM}, taken into account in the model. When measuring $\Ya$, conditioning does not distort the Rabi oscillation. This is because the kick-back of $\Ma=\pm1$ is a $z$ rotation of $\Q$ by $\pm \epsilon/2$, leading to the same $x$ projection.  Conditioning on a $\Za$ measurement produces the most striking difference: while $\Ma=+1$ imposes no kick-back, $\Ma=-1$ imposes a $z$ rotation of $\pi$. Ideally, both curves are unit-amplitude sinusoids with opposite phase, independent of $\epsilon$. However, for $\epsilon=45^\circ$, the $\Ma =-1$ set is dominated by false negatives. As $\epsilon$ increases, true $\Ma=-1$ counts become more abundant and we  observe the expected sign reversal in the conditioned curve with $\epsilon=90^\circ$. Note that despite the difference in conditioned curves for the different $\A$ measurement bases, the three unconditioned curves are nearly identical (See~\cite{SOM} for more values of $\epsilon$). This is consistent with the expectation that measurement induced dephasing is independent of the ancilla measurement basis~\cite{Wiseman09}.

As a benchmark of the complete indirect-measurement scheme, we extract quantum efficiencies $\eta_{i}$  characterizing the loss of quantum information~\cite{Korotkov08} for measurement outcome $\Ma=i$:
\[
\eta_{i}\equiv \left(\frac{\avgb{\Xq'}^2+\avgb{\Yq'}^2}{1-\avgb{\Zq'}^2}\right)^{1/2}_{|
\Ma=i}\bigg/\left(\frac{\avgb{\Xq}^2+\avgb{\Yq}^2}{1-\avgb{\Zq}^2}\right)^{1/2}.
\]
Loss originates in the single-shot readout infidelity of $A$, the residual excitation in $A$ and $B$, and decoherence of $Q$, $A$, and $B$ during the interaction. Without decoherence, $\eta_{i}$ would be independent of input qubit state (see~\cite{SOM} for full expressions).  Using the calibrated infidelities and residual excitations, we estimate $\eta_{\pm1}=0.94~(0.94)$ at $\epsilon=45^\circ$ and $0.85~(0.83)$ at $90^\circ$. Including decoherence and averaging over the qubit Bloch sphere, we estimate $\bar{\eta}_{\pm1}=0.77~(0.71)$ at $\epsilon=45^\circ$ and $0.69~(0.60)$ at $90^\circ$.

\begin{figure}
\includegraphics[width=\columnwidth]{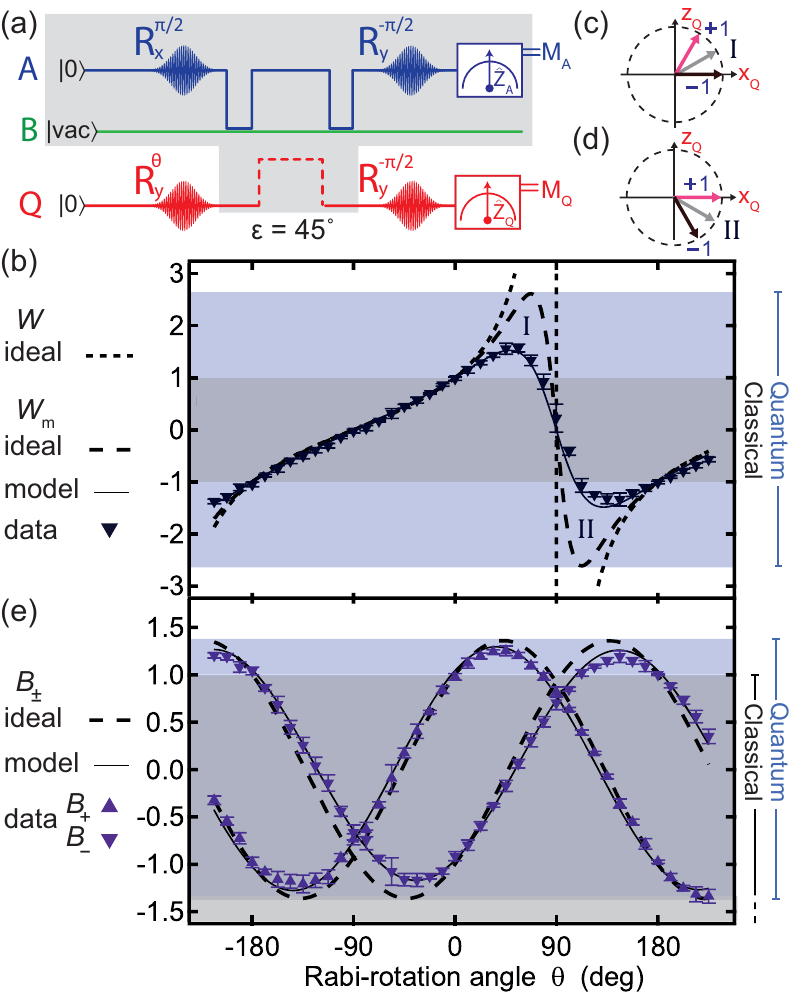}
\caption{(color online). Observation of non-classical weak values and Leggett-Garg inequality violations (measurement strength $\epsilon=45^\circ$). (a) Pulse sequence. Note that qubit roles are swapped ($\Q=\Qtwo$  and $\A=\Qone$) compared to previous figures in order to minimize errors when conditioning on $\Mq$. (b) Polar-angle dependence of modified weak value $\Wm$ (see text for definition and normalization procedure). The MAR bound $|\Wm|<1$ is amply exceeded. From the quantum perspective, the extrema in $\Wm$ at I and II can be understood using the Bloch spheres in (c) and (d), respectively.  Ideally, at I (II), the $\Ma=-1(+1)$ kick-back aligns $\Q$ with the $+\xq$ axis, perfectly correlating $\Mq=-1$ with $\Ma=+1(-1)$. (e) Measured averaged Leggett-Garg operators $\Bpm$ defined in text. One of the inequalities $-3\leq\Bpm\leq1$ is violated whenever non-classical $\Wm$ is observed.}
\end{figure}

Finally, we combine the abilities to perform partial and projective measurements to observe non-classical weak values, detect LGI violations, and demonstrate their correspondence~\cite{Williams08, Goggin11} (Fig.~4). The partial measurement of $\Q=\Qtwo$ is performed via $\A=\Qone$ and $\Hone$ and the projective measurement via $\Htwo$. We measure the partial-measurement average conditioned on strong-measurement result $\Mq=-1$,
\[
\Wm \equiv \avgb{\Manorm}_{|\Mq=-1},
\]
where $\Manorm \equiv \left(\Ma - m_{\mathrm{off}}\right)/m_{\mathrm{pk}}$ is the offset and rescaled partial-measurement result so that $\avgb{\Manorm}=\pm1$ for $\ketq{0} (\ketq{1})$. MAR constrains $|\Wm| \leq 1$, while quantum mechanics allows $|\Wm| \leq 1/\sin(\epsilon/2)$. We call $\Wm$ a modified weak value because it differs in the ideal quantum setting (perfect interaction and measurements) from the standard definition~\cite{Aharonov88} of the weak value $\W$ of operator $\Zq$ between initial state $\ket{\theta_Q}$ and final state $\ket{-\pi/2_Q}$,
\[
\W \equiv \frac{\bra{-\pi/2_Q} \Zq \ket{\theta_Q}}{\braket{-\pi/2_Q}{\theta_Q}}.
\]
Specifically, the digital character of ancilla-based measurement~\cite{Wu09,SOM} regularizes $\Wm$ near $\theta=\pi/2$, where $W$ diverges [Fig.~4(b)].
In parallel, we consider the averaged Leggett-Garg operators
\[
\Bpm \equiv  \pm \avgb{\Manorm} \mp \avgb{\Manorm\Mq} + \avg{\Mq}.
\]
Under MAR, a weak, non-invasive measurement and subsequent strong measurement of a two-level system satisfy the LGIs $-3\leq \Bpm \leq 1$. For the sequence in Fig.~4(a), quantum mechanics allows $|\Bpm| \leq \sqrt{(\cos(\epsilon)+3)/2}$.

We capture the smooth crossing of MAR bounds for $\Wm$ and $\Bpm$ by performing the experiment in Fig.~4(a) over a range of initial qubit states $\ket{\theta_Q}$. We observe a maximum $\Wm$ of $1.57\pm0.08$. In turn, the averaged Leggett-Garg operators $\Bp$ and $\Bm$ peak at $1.25\pm0.05$ and $1.19\pm0.06$, respectively. The data clearly show that one of the two LGIs is violated whenever $\Wm$ is non-classical. This correspondence, predicted in Ref.~\onlinecite{Williams08} and previously only demonstrated with photons~\cite{Goggin11}, becomes the more interesting upon noting that $B_{\pm}$ averages all measurements while $\Wm$ uses only the post-selected fraction for which $\Mq=-1$.

In conclusion, we have realized an indirect measurement of a transmon qubit with high quantum efficiency and tunable measurement strength. Our scheme consists of a partially entangling interaction between the qubit and an ancilla, followed by projective ancilla measurement using a dedicated, dispersively-coupled resonator. We have measured the kick-back of such measurements on the qubit as a function of interaction strength and ancilla measurement basis, finding close agreement with theory. Non-classical weak values are observed upon conditioning ancilla measurements on the outcome of a projective measurement of the qubit. Their  predicted correspondence with LGI violations is demonstrated for the first time in a solid-state system.
The combination of high-quality factor bus, individual readout resonators and feedline here demonstrated constitutes a scalable architecture~\cite{QSpeedup} with frequency-multiplexable single-qubit control and readout~\cite{Chu12}. Future experiments using this architecture will target the realization of an ancilla-based 4-qubit parity measurement as needed for surface-code quantum error correction~\cite{Fowler12}.

\begin{acknowledgments}
We thank D.~Thoen and T.~M.~Klapwijk for NbTiN thin films, A.~Frisk Kockum for contributions to theoretical modeling, and S.~Ashhab, Y.~Blanter, M.~H.~Devoret, M.~Dukalski, G.~Haack, and R.~Hanson for helpful discussions. We acknowledge funding from the Dutch Organization for Fundamental
Research on Matter (FOM), the Netherlands Organization for Scientific Research (NWO, VIDI scheme), the EU FP7 project SOLID, and the Swedish Research Council.
\end{acknowledgments}

\end{document}


\title{Supplementary material for ``Partial-measurement back-action and non-classical weak values in a superconducting circuit"}
\author{J.~P.~Groen}
\author{D.~Rist\`e}
\affiliation{Kavli Institute of Nanoscience, Delft University of Technology, P.O. Box 5046,
2600 GA Delft, The Netherlands}
\author{L.~Tornberg}
\affiliation{Microtechnology and Nanoscience, MC2, Chalmers University of Technology, SE-412 96 Goteborg, Sweden}
\author{J.~Cramer}
\affiliation{Kavli Institute of Nanoscience, Delft University of Technology, P.O. Box 5046,
2600 GA Delft, The Netherlands}
\author{P.~C.~de Groot}
\affiliation{Kavli Institute of Nanoscience, Delft University of Technology, P.O. Box 5046,
2600 GA Delft, The Netherlands}
\affiliation{Max Planck Institute for Quantum Optics, Garching 85748, Munich, Germany}
\author{T.~Picot}
\affiliation{Kavli Institute of Nanoscience, Delft University of Technology, P.O. Box 5046,
2600 GA Delft, The Netherlands}
\affiliation{Laboratory of Solid-State Physics and Magnetism, KU Leuven, Celestijnenlaan 200D, 3001 Leuven, Belgium}
\author{G.~Johansson}
\affiliation{Microtechnology and Nanoscience, MC2, Chalmers University of Technology, SE-412 96 Goteborg, Sweden}
\author{L.~DiCarlo}
\affiliation{Kavli Institute of Nanoscience, Delft University of Technology, P.O. Box 5046,
2600 GA Delft, The Netherlands}
\date{\today}

\maketitle

\section{Device fabrication}
The two-qubit, three-resonator chip was fabricated on a sapphire substrate ($430~\um$ thick, C-Plane). Following in-situ cleaning of the substrate in Ar for $2~\mins$, a $65~\nm$ thick niobium titanium nitride (NbTiN) film~\cite{Barends10} was DC-sputtered. Superconducting coplanar waveguide structures ($10~\um$ central conductor width, $4.2~\um$ gaps) were then defined using a negative electron-beam resist (SAL-601) and reactive-ion etching in a $\mathrm{SF}_6/\mathrm{O}_2$ plasma. Finally, the two transmons were patterned by standard electron-beam lithography and Al double-angle evaporation ($20~\nm$ bottom and 70$~\nm$ top layer thicknesses, with intermediate oxidation for $10~\mins$ at $0.55~\mathrm{mBar}$). During fabrication, the device was exposed three times to an $\mathrm{O}_2$ plasma to remove organic residues: before evaporation, after lift-off, and after dicing the sample to $2~\mm \times 7~\mm$.

\section{Experimental methods}
\begin{figure*}
\includegraphics[width=1.5\columnwidth]{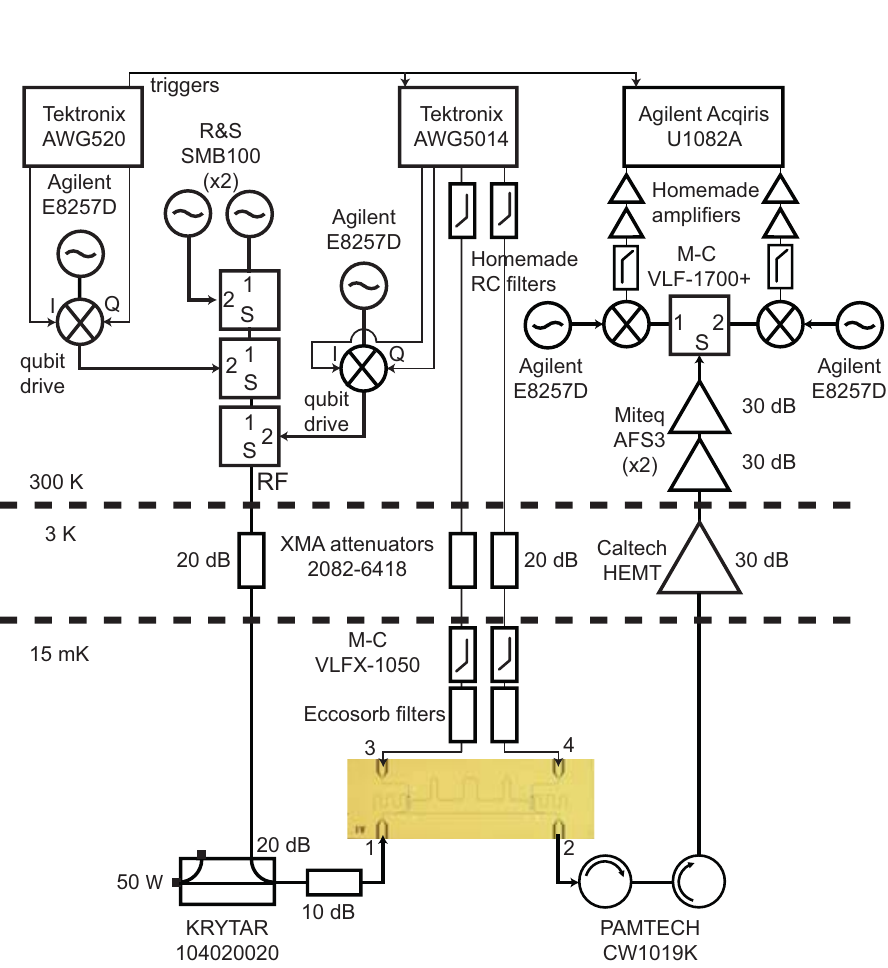}
\caption{Experimental setup. Arbitrary waveform generators Tektronix AWG520 and AWG5014, with 10- and 14-bit resolution, respectively,
and $1~\ns$ sampling rate produce voltages directly applied to the flux-bias lines, the single-sideband $I$-$Q$
modulation envelopes for the microwave tones driving single-qubit rotations, and the pulse envelopes for measurement tones. Flux pulses are conditioned by a series combination of DC block, attenuation, $LC$ low-pass filter and homemade coaxial eccosorb filter (based on Ref.~\onlinecite{Santavicca08}) before reaching ports 3 and 4. All measurement and qubit-drive pulses are combined at room temperature. Inside the dilution refrigerator, they are coupled to the feedline input (port 1) following $50~\dB$ attenuation. On the feedline output (port 4), an amplification chain with $\sim100~\dB$ gain, two I-Q mixers and a two-channel averaging digitizer ($1~\ns$, 8-bit sampling) process the two readout signals.}
\label{Fig:Setup}
\end{figure*}

\begin{figure*}
\includegraphics[width=1.4\columnwidth]{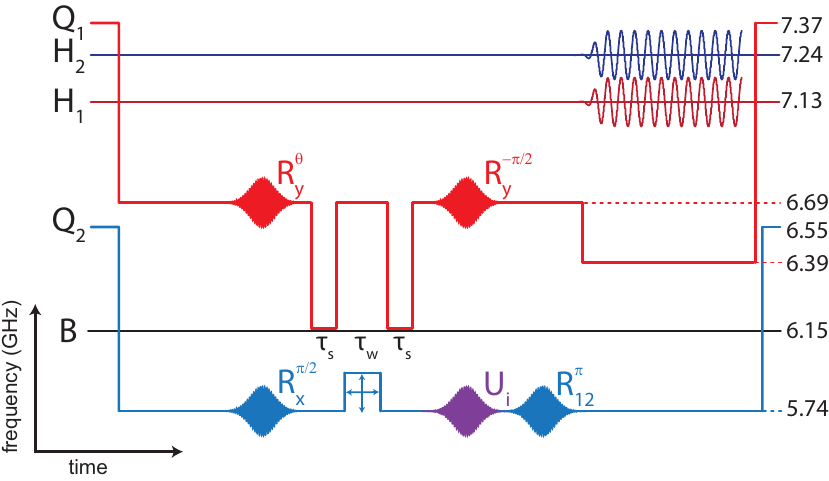}
\caption{Detailed sequence of microwave and flux pulses realizing the indirect measurement in Fig.~2(a). The qubits are detuned from each other to minimize crosstalk in qubit control and readout. $\Qone=\Q$ ($\Qtwo=\A$) is pulsed to the bias point $[6.69~(5.74)~\GHz]$ where all single-qubit operations are performed.  All single-qubit gates are realized with resonant DRAG~\cite{Chow10b} pulses, with standard deviation $\sigma = 4~\ns$, and $\pm2\sigma$ truncation.  The rotation axis is set using I-Q (vector) modulation (see Fig.~\ref{Fig:Setup}). The controlled-$z$ rotation is realized by coherently swapping the $\Qone$ state into the bus $\B$ in $\tswap=7~\ns$, waiting a calibrated time $\twait$ (Fig.~\ref{Fig:Gate}), and then swapping the $\Bus$ state back onto $\Qone$. The photon-number dependent shift of  $\Qtwo$ during $\twait$ produces the two-qubit phase that can partially entangle $\Qone$ and $\Qtwo$. The measurement pre-rotations $U_i$ for ancilla measurement in the $\Xa$, $\Ya$, and $\Za$ bases are $R_{\ya}(-\pi/2)$, identity, and $R_{\xa}(-\pi/2)$, respectively. Before measurement, $\Qone$ is flux-pulsed to $~6.39~\GHz$ and a $\pi$ pulse on the 1-2 transition of $\Qtwo$ is applied to maximize fidelity. A $2~\ns$ buffer is inserted between adjacent pulses to avoid any overlap.}
\label{Fig:Pulses}
\end{figure*}

\begin{figure*}
\includegraphics[width=1.6\columnwidth]{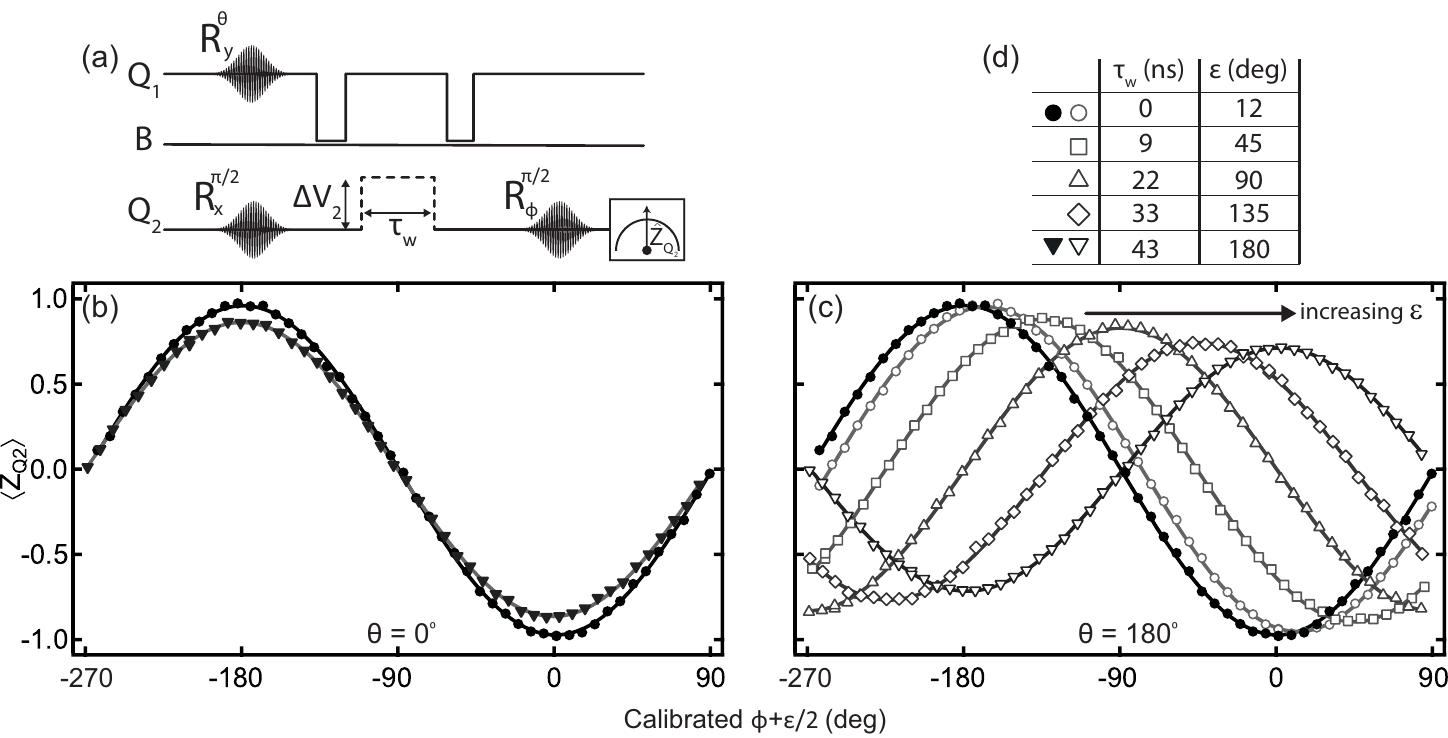}
\caption{
Calibration of the partial-measurement interaction step. (a) Pulse scheme for measuring the single-qubit phase of $\Qtwo$ and the two-qubit phase acquired during flux pulsing. We
measure the shift in the azimuthal phase of the $\Qtwo$ Bloch vector with $\Qone$ prepared nominally in $\ket{0}$ ($\theta=0$) and in $\ket{1}$ ($\theta=180^\circ$). The peak of the $\theta=0~(180^\circ)$ curve in (b) [(c)] is matched to azimuthal angle $\phi=-180^\circ\mp\epsilon/2$. In this way  we implement virtual $z$-gate corrections~\cite{Reed12} as common in nuclear-magnetic-resonance experiments~\cite{Steffen00}.
(d) Table of calibrated $\epsilon$ for various conditions of the $\Qtwo$ flux pulse in between the $\Qone$-$\Bus$ swaps. The duration $\twait$ ($1~\ns$ resolution) and the flux-pulse amplitude $\Delta V_2$ are used to coarse- and fine-tune $\epsilon$, respectively.  The single-qubit phase acquired by $\Qone$ is calibrated by interchanging the rotations applied to $\Qone$ and $\Qtwo$, and is similarly compensated.
}
\label{Fig:Gate}
\end{figure*}

\begin{figure*}
\includegraphics[width=1\columnwidth]{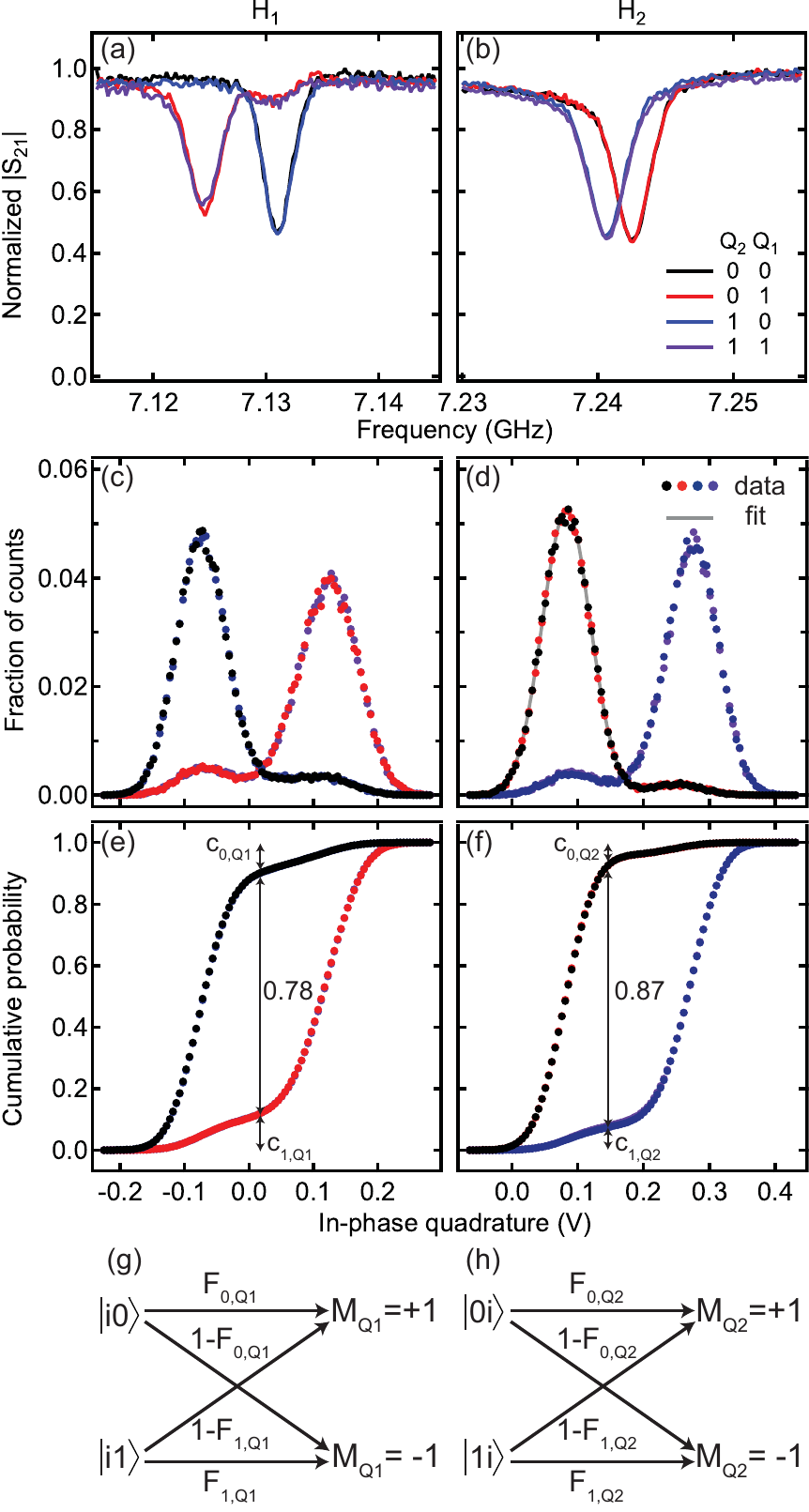}
\caption{Characterization of individual qubit readouts. (a,b) Averaged pulsed measurement of feedline transmission as a function of frequency near the fundamental-mode (quarter-wave) frequencies of (a) $\Hone$ and (b) $\Htwo$ immediately following preparation of $\ket{ji}$ with $\pi$ pulses ($j,i\in\{0,1\}$ denote the state of $\Qtwo$ and $\Qone$, respectively). Here, we use $280~\ns$ integration time and $-121~\dBm~(-111~\dBm)$ input power at port~ 1. Evidently, $\Hone$ ($\Htwo$) is predominantly sensitive to the state of $\Qone$ ($\Qtwo$), with $-6.4~\MHz$ ($-2.0~\MHz$) dispersive shift. (c,d) Single-shot histograms at $f_{\Hone}=7.130~\GHz$ and $f_{\Htwo}= 7.243~\GHz$ [$600~\ns$ integration time, $-110~\dBm$ and $-111~\dBm$ input power, respectively]. Using the best fit of a double Gaussian to the $\Htwo$ histograms for $\Qtwo$ nominally in $\ket{0}$, we estimate that $4\%$ of counts fall in the peak corresponding to $\ket{1}$. We attribute this fraction to residual excitation of $\Qtwo$. This hypothesis is supported by other measurements (not shown) at variable power and duration, giving similar fit results. The power dependence observed in histograms for $\Hone$ with $\Qone$ nominally in $\ket{0}$, however, does not allow this analysis. When modeling, we thus assume a residual excitation of $\Qone$ equal to that measured for $\Qtwo$. (e,f) Cumulative probability of histograms in (c,d). The readout contrast is $1-c_{0}-c_{1}=0.779\pm 0.005$ for $\Hone$ and $0.870\pm0.007\%$ for $\Htwo$. (g,h) Readout error model for $\Qone$ and $\Qtwo$. Accounting for the contrast reduction induced by residual qubit excitation and assuming perfect pulses, we extract single-shot readout fidelities $F_{0,Q1}= 0.93$, $F_{1,Q1}=  0.92$, $F_{0,Q2}= 0.99$, and $F_{1,Q2}= 0.95$.}
\label{Fig:Readout}
\end{figure*}

This section consists of four fully-captioned figures providing further detail on experimental methods. The complete setup, both inside and outside the dilution refrigerator, is illustrated in Fig.~\ref{Fig:Setup}. The
detailed flux-pulsing scheme showing the bias points chosen for single-qubit control, readout and interactions with the bus is shown in Fig.~\ref{Fig:Pulses}.  Figure~\ref{Fig:Gate} presents the calibration of the interaction step for the values of measurement strength $\epsilon$ used. Finally, Fig.~\ref{Fig:Readout} demonstrates the use of resonators $\Hone$ and $\Htwo$ for individual readout of $\Qone$ and $\Qtwo$, respectively, and the extraction of single-shot readout fidelities and residual qubit excitations.

\section{Extended results}
\begin{figure*}
\includegraphics[width=1.2\columnwidth]{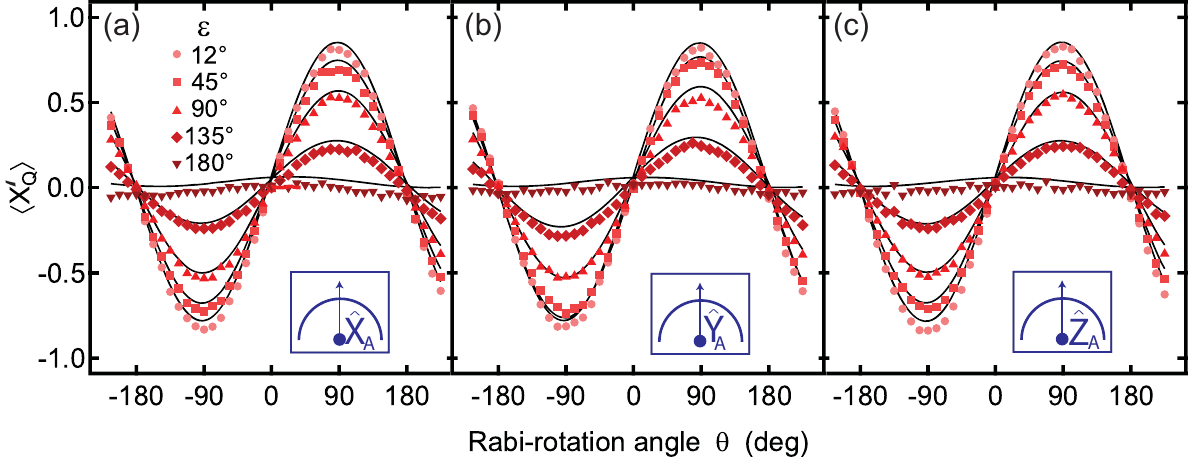}
\caption{$\avg{\Xq'}$ as a function of the qubit Rabi-rotation angle $\theta$, for three choices of basis for the ancilla measurement $\Ma$: (a) $\Xa$, (b) $\Ya$, (c) $\Za$. Panel (a) replicates the data in Fig.~2(e). Clearly, unconditioned measurements of $\Q$ do not depend on the choice of ancilla measurement basis.}
\label{Fig:MeasDephasing}
\end{figure*}

\begin{figure*}
\includegraphics[width=2\columnwidth]{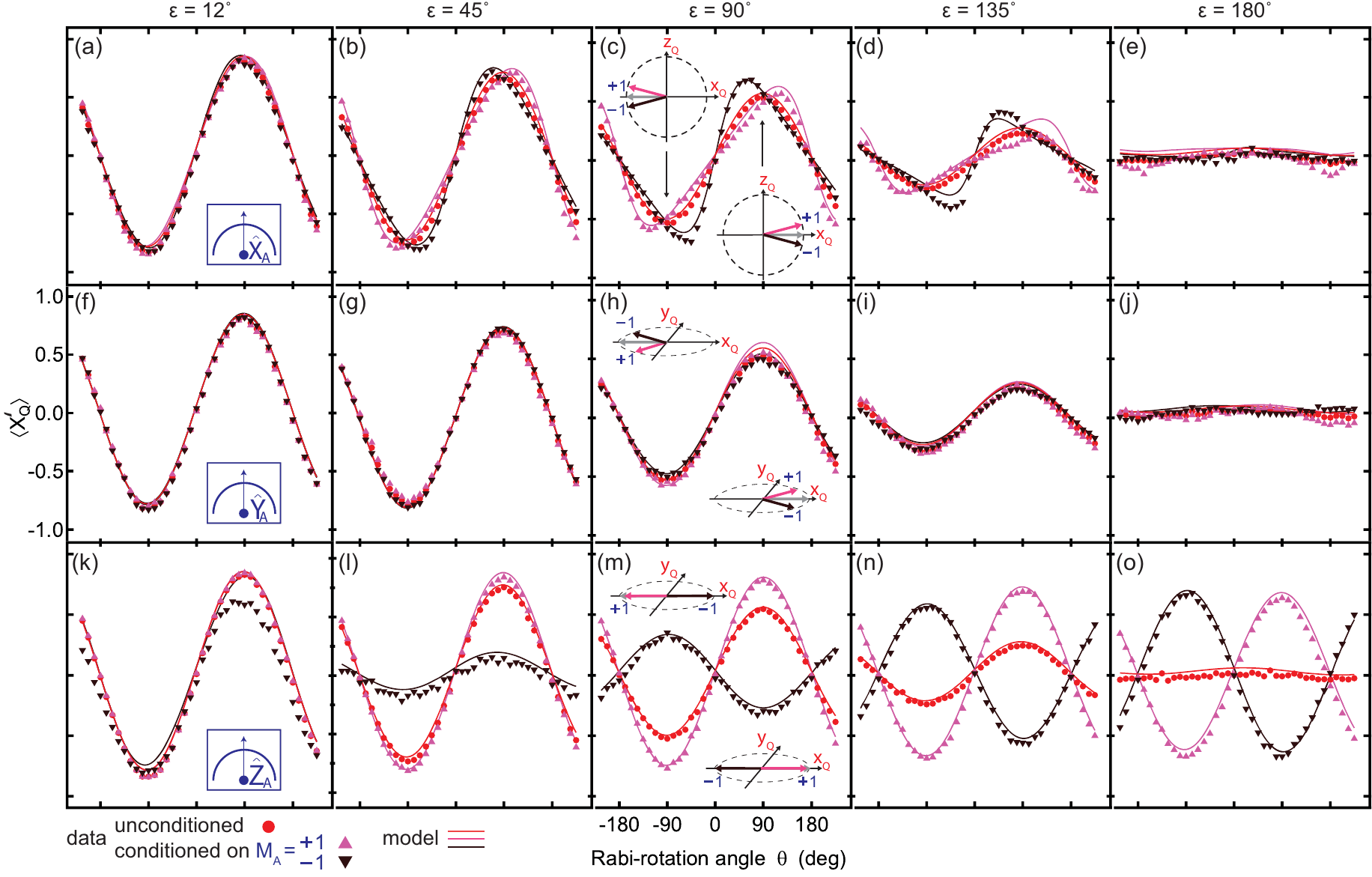}
\caption{Measurement back-action over a wide range of measurement strength $\epsilon$.
Panels (b), (c), (g), (h), (l), and (m) replicate the data in Fig.~3.}
\label{Fig:BackAction}
\end{figure*}

\begin{figure*}
\includegraphics[width=2\columnwidth]{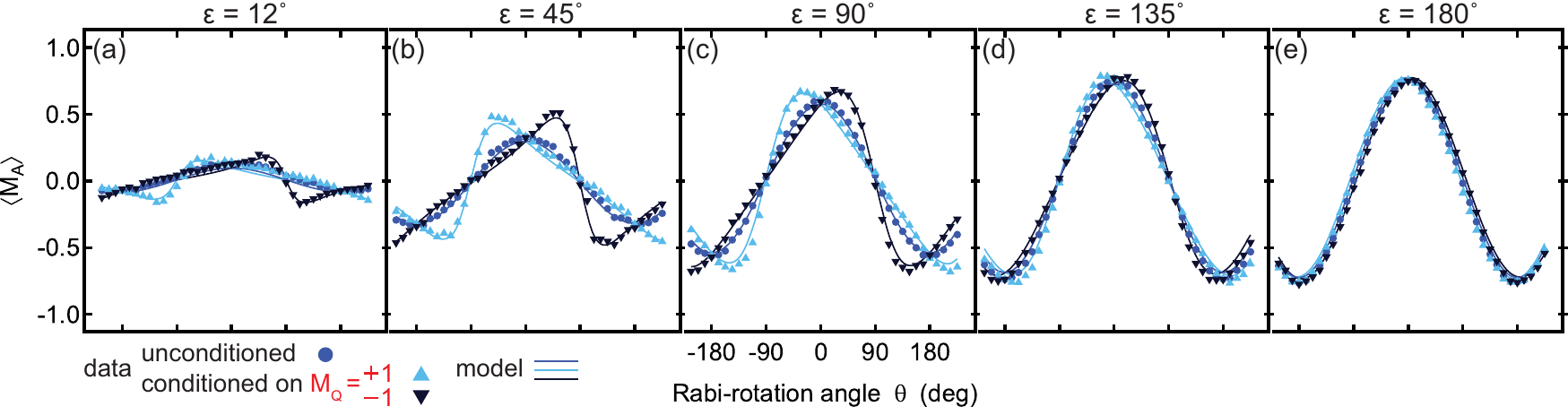}
\caption{Weak-value measurements over a wide range of measurement strength $\epsilon$.
Panel (b) shows the raw data used in Fig.~4(b).}
\label{Fig:WeakValues}
\end{figure*}

\begin{figure*}
\includegraphics[width=1.1\columnwidth]{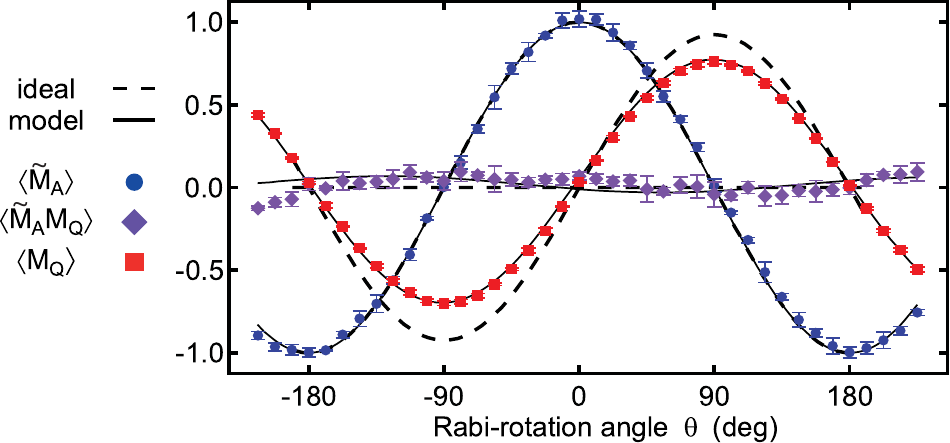}
\caption{The three terms $\avgb{\Manorm}$, $\avgb{\Manorm\Mq}$, and $\avg{\Mq}$ contributing to  the averaged Leggett-Garg operators $\Bpm$ shown in Fig.~4. As discussed in the main text,  $\Manorm$ is the partial-measurement result, offset and rescaled so that $\avgb{\Manorm} = \pm 1$ for $\ketq{0} (\ketq{1})$). In the ideal quantum setting, the two-qubit correlation term $\avgb{\Manorm\Mq} = 0$ for all $\theta$ and the reduced contrast in $\avg{\Mq}$ arising from partial-measurement kick-back is $\avg{\Mq}=\sin(\theta)\cos(\epsilon/2)$. The vertical asymmetry in $\avg{\Mq}$ observed in data and model is due to asymmetric errors in the readout of $\Q$~(Fig.~\ref{Fig:Readout}).}
\label{Fig:LGIterms}
\end{figure*}

This section consists of four fully-captioned figures extending the results and backing claims in the main text. Figure~\ref{Fig:MeasDephasing} demonstrates that partial-measurement induced dephasing is independent of the basis chosen for the ancilla measurement. Figure~\ref{Fig:BackAction} shows the measurement-induced back-action at more values of measurement strength $\epsilon$ than Fig.~3. Figure~\ref{Fig:WeakValues} shows the raw data contributing
to the measurement of weak values, complementing Fig.~4(b) and showing raw measurements at more values of $\epsilon$. Figure~\ref{Fig:LGIterms} shows the three terms contributing to the averaged Leggett-Garg operators $\Bpm$, complementing Fig.~4(c).

\section{Theory}
\subsection{Hamiltonian model}
We consider a system of two transmons ($\Qone$ and $\Qtwo$) coupled to one bus resonator. The system Hamiltonian is
\be\label{Eq:H}
H =  H_B +  \sum_{i = 1}^2 \left( H_{Qi} + I_{Qi} + D_{Qi}(t)\right),
\ee
where $H_B$ and $H_{Qi}$ describe the non-interacting dynamics of the bus and qubit $Q_i$, respectively~\cite{Koch07}:
\be
H_B = \hbar\omega_{B} \dagg{a}a, \qquad H_{Q_i} = \sum_j \hbar \omega_{j,Qi} \ket{j_{Qi}}\bra{j_{Qi}}.
\ee
Here, $\omega_B$ is the bus resonance frequency and $\hbar\omega_{j,Qi}$ is the energy of the $j^{\mathrm{th}}$ level of qubit $Q_i$. In the transmon regime $\EJ/\EC \gg 1$ valid here~\cite{Koch07}, the interaction between a transmon and the bus can be modeled by an extended Jaynes-Cummings-type coupling~\cite{Koch07}:
\be
I_{Qi} = \hbar \sum_j g_{j+1,j}\left( \ket{j+1_{Qi}}\bra{j_{Qi}}a + \mathrm{h.c.} \right),
\ee
with coupling strengths $g_{j+1,j} = \sqrt{j+1} g_0$, where $2 g_0$ is the vacuum Rabi splitting. The driving Hamiltonian is
\be
\begin{array}{lcl}
D_{Qi}(t) &=\sum_{i,j,d} &\left( \epsilon_d(t)\sqrt{j+1}\ket{j+1_{Qi}}\bra{j_{Qi}} e^{-i \omega_d t+\phi_d}\right.\\
&& \ \left. +\ \mathrm{h.c.} \right),
\end{array}
\ee
where $\epsilon_d(t)$, $\phi_d$, and $\omega_d$ denote the amplitude, phase, and frequency of drive $d$, respectively. When each qubit is far detuned from the bus, i.e.  $g_{j+1,j}/| \omega_{j+1,Qi} - \omega_{j,Qi} - \omega_B | \ll1$, the system is described by a dispersive Hamiltonian~\cite{Koch07} with transmon and transmon-bus interaction terms
\bea\label{Eq:dispersive}
H_{Qi}^\text{d} &=& \hbar \sum_j \omega_{j,Qi} \ket{j_{Qi}}\bra{j_{Qi}} + \chi_{j,Qi} \ket{j+1_{Qi}}\bra{j+1_{Qi}}, \nonumber \\
I_{Qi}^\text{d} &=&  -\hbar \chi_{0,Qi} \dagg{a}a \ket{0_{Qi}}\bra{0_{Qi}}  \nonumber\\
&+& \hbar \sum_{j\geq1}  (\chi_{j-1,Qi} - \chi_{j,Qi} )\dagg{a}a \ket{j_{Qi}}\bra{j_{Qi}}.
\eea
Here, $\chi_{j,Qi} = g^2_{j+1,j}/(\omega_{j+1,Qi} - \omega_{j,Qi} - \omega_B)$ is the dispersive bus-transmon coupling. The dispersive regime allows for an analytical solution for the shift of all resonance frequencies, and simplifies the modeling of the gate sequence. We therefore use Eq.~(\ref{Eq:dispersive}) to simulate the dynamics in the off-resonant passages with $H_{Qi} \to H_{Qi}^\text{d}$ and  $I_{Qi} \to I_{Qi}^\text{d}$ in Eq.~(\ref{Eq:H}).

Apart from the coherent dynamics, the coupling of the
system to the environment leads to dissipative evolution. Assuming weak coupling, the Markovian master equation describing the system evolution is~\cite{thesisBishop10}
\bea\label{Eq:master}
\dot{\rho} &=& -i[H,\rho] + \kappa\lind{a}\rho+\kappa'\lind{\dagg{a}}\rho\\
&+&\sum_{i=1}^2 \gamma_{01,Qi}\lind{\sigma^{-}_{Qi}}\rho \nonumber \\
&+&\sum_{i=1}^2 \gamma_{10,Qi}\lind{\sigma^{+}_{Qi}}\rho \nonumber \\
&+&\sum_{i=1}^2 \frac{\gamma_{\phi,Qi}}{2}\lind{Z_{Qi}}\rho \equiv \mathcal{L}\rho \nonumber,
\eea
where $\lind{L}\rho=(2 L \rho \dagg{L}-\dagg{L} L\rho-\rho L \dagg{L})/2$,
$\kappa~(\kappa')$ is the bus photon damping (excitation) rate, and
$\gamma_{01,Qi}$, $\gamma_{10,Qi}$, and $\gamma_{\phi,Qi}$ are relaxation, excitation, and pure dephasing rates for $Q_i$. Here, we have explicitly truncated the Hilbert space of each transmon to the lowest two levels. Equation~(\ref{Eq:master}) is conveniently solved in Liouville space, where $\rho$ is a vector. The solution is given by $\rho(t) = e^{\mathcal{L}t}\rho(0)$, which can be solved by numerically diagonalizing the propagator $\mathcal{L}$. In the simulation, we truncate the bus Hilbert space to $n =\{0,1,2\}$ photons.

The model is fit to the data with the following fit parameters: the frequency of $\Qone$ during the interaction with $\B$, the two single-qubit phases, and the amplitude error in the pre-measurement rotation for $\Q$. These parameters are independently fit for each pair of simultaneous measurements of $\Qone$ and $\Qtwo$.  For conditioned measurements, we also fit the readout fidelities of the conditioning readout. Fixed model parameters include $4\%$ excitation in each element [measured for $\Qtwo$ (Fig.~\ref{Fig:Readout}) and assumed equal for $\Qone$ and $\B$],  the measured energy-equilibration times $1.4$, $2.5$, and $5.3~\us$ for $\Qone$, $\Qtwo$, and $\B$, respectively, and pure-dephasing time $1.6$~$(1.8)~\us$ for $\Qone$~$(\Qtwo)$. Typical best-fit values give an absolute error within $5^\circ$ for $\epsilon$, $10^\circ (3^\circ)$ for the phase of $\Qone$~$(\Qtwo)$, and $2\%$ for the rotation amplitude. 

\subsection{Partial measurement : ideal case}
For reference, we analyze the partial-measurement scheme with perfect interaction and measurement steps.
In the interaction step, the qubit and ancilla evolve as
\[
\ketq{\Psi}\keta{0} \rightarrow \  U \ketq{\Psi}\keta{0},
\]
with
\[
U=\cos(\epsilon/4) \Iq \Ia -i \sin(\epsilon/4)\Zq \Ya.
\]
Upon performing a projective measurement of the ancilla with operator $O_A = \sum_i \lambda_i \keta{i}\braa{i}$, the (unnormalized) post-measurement qubit state for result $\Ma=\lambda_i$ is
\[
\ketq{\Psi'}=\Omega_{\lambda_i} \ketq{\Psi},
\]
where $\Omega_{\lambda_i}= \braa{i} U \keta{0}$.
The probability of getting measurement result $\Ma=\lambda_i$ is
\[
P_{\Ma=\lambda_i}=\braq{\Psi} \Omega_{\lambda_i}^\dagger \Omega_{\lambda_i} \ketq{\Psi}.
\]
When the measurement results are disregarded, the post-measurement qubit density matrix is
\begin{align*}
\rhoq'&=\sum_i \Omega_{\lambda_i} \rhoq \Omega_{\lambda_i}^\dagger\\
&=\tr_{A}\left[U \rhoq \keta{0}\braa{0} U^\dagger \right]\\
&=\cos^2(\eps/4)\rhoq +\sin^2(\eps/4)\Zq \rhoq \Zq,\\
\end{align*}
where $\rhoq=\ketq{\Psi}\braq{\Psi}$ is the initial qubit density matrix. Clearly, $\rhoq'$ is independent of the choice of ancilla measurement basis. It is straightforward to show that the transformation
\[
\left(\begin{array}{c}
\avg{\Xq}\\
\avg{\Yq}\\
\avg{\Zq}
\end{array}\right)
\rightarrow
\left(\begin{array}{c}
\avg{\Xq'}\\
\avg{\Yq'}\\
\avg{\Zq'}
\end{array}\right)
\]
of the qubit Bloch vector is
\[
\left(\begin{array}{c}
\avg{\Xq'}\\
\avg{\Yq'}\\
\avg{\Zq'}
\end{array}\right)
=
\left(\begin{array}{c}
\cos(\eps/2)\avg{\Xq}\\
\cos(\eps/2)\avg{\Yq}\\
\avg{\Zq}
\end{array}\right)
\]

\subsubsection{Ancilla measurement in $\Xa$}
The operation elements evaluate to
\[
\Omega_{\pm1} =\frac{1}{\sqrt{2}}\cos(\eps/4)\Iq \pm \frac{1}{\sqrt{2}}\sin(\eps/4) \Zq,
\]
giving the measurement result probabilities
\[
P_{\Ma = \pm 1} =\frac{1}{2}\left(1 \pm \sin(\eps/2)\la \Zq \ra \right)\\
\]
and the measurement average
\[
\avg{\Ma} = \avg{\Zq}\sin(\eps/2).
\]

We briefly visualize how these operation elements kick a qubit initially in the pure state $\ketq{\psi}=\ket{\theta,\phi}=\cos(\theta/2)\ketq{0}+e^{i\phi}\sin(\theta/2)\ketq{1}$.
Because these operation elements are real-valued and diagonal, the post-measurement state $\ketq{\psi'}=\ket{\theta',\phi'}$ has the
same azimuthal angle, $\phi'=\phi$. The polar angle transforms as
\[
\tan\left(\theta'/2\right)=\tan(\theta/2)\frac{\cos(\eps/4)-\Ma\sin(\eps/4)}{\cos(\eps/4)+\Ma\sin(\eps/4)}.
\]
We consider a few special cases. For a qubit initially on the equator of the Bloch sphere ($\theta=\pi/2$), a positive (negative) measurement
kicks the qubit toward the north (south) pole,  decreasing (increasing) its polar angle by $\eps/2$. A qubit initially at one of the poles $(\theta=0,\pi)$ remains at the pole regardless of measurement result.
For all other cases, the update formula shows that the change in polar angle is not equal in magnitude. In the northern (southern) hemisphere, a positive result decreases the polar angle less (more) than a negative result
increases it.

\subsubsection{Measurement in $\Ya$}
The operation elements are
\[
\begin{array}{lcl}
\Omega_{\pm 1} &=& \frac{1}{\sqrt{2}}\left(\cos(\epsilon/4) \mp i \sin(\epsilon/4) \Zq \right)\\
&=&\frac{1}{\sqrt{2}}R_{\zq}(\pm \epsilon/2),
\end{array}
\]
yielding
\[
P_{\Ma = \pm 1}=\frac{1}{2},\quad \avg{\Ma}=0.
\]
The kick-back of  $\Ma=\pm1$ is evidently a $z$-axis rotation of $\Q$ by $\pm\epsilon/2$. This kick-back is independent of the initial qubit state.

\subsubsection{Measurement in $\Za$}
The operation elements are
\begin{align*}
\Omega_{+1}&=\cos(\eps/4) \Iq\\
\Omega_{-1}&=\sin(\eps/4) \Zq,
\end{align*}
yielding
\begin{align*}
P_{\Ma =+1}&=\cos^2(\eps/4)\\
P_{\Ma = -1}&=\sin^2(\eps/4),
\end{align*}
and thus a measurement average
\[
\avg{\Ma}=\cos(\eps/2).
\]
For the most likely result $(\Ma=+1)$, there is no back-action. The rare result $(\Ma=-1)$ rotates the qubit by $\pi$ around the $z$ axis.
\vspace{.5cm}
\subsection{Modified weak value: ideal case}
Consider performing the general partial plus projective measurement scheme (Fig.~4) starting with the qubit in $\ketq{\psi}=\ket{\theta}$ and the ancilla in $\keta{0}$.
A projective measurement of $\Xq$ on $\Q$ following the interaction step has for operation elements on $\A$
\[
\Omega_{\Mq = \pm1} = \braq{\pm\pi/2} U \ketq{\theta}\keta{0}.
\]
The (unnormalized) post-measurement ancilla states are
\begin{align*}
\keta{\psi'}=&\Omega_{\Mq = \pm1} \keta{0}\\
=&\cos(\eps/4)\braket{\pm\pi/2_{Q}}{\theta_{Q}}\keta{0}\\
&+\sin(\eps/4)\bra{\pm\pi/2_{Q}}\Zq\ket{\theta_{Q}}\keta{1}.
\end{align*}
From these we can calculate the expectation value of $\Ma$ ($\Xa$ basis) conditioned on $\Mq=\pm1$:
\begin{align*}
\avg{\Ma}_{|\Mq=\pm1}=&\frac{\sin(\eps/2)}{ \cos^2(\eps/4)+\sin^2(\eps/4)|W(\Zq,\pm\pi/2,\theta)|^2}\times\\
&\mathrm{Re}\{W(\Zq,\pm \pi/2,\theta)\},
\end{align*}
where
\[
W(A_Q,\theta ',\theta)\equiv \frac{\braq{\theta'} A_Q \ketq{\theta}}{\braket{\theta'_{Q}}{\theta_{Q}}}
\]
is the weak value of Hermitian operator $A_Q$ between initial state $\ketq{\theta}$ and final state $\ketq{\theta'}$.

\subsection{Measurement model}
In order to include the readout errors in the model curves, we numerically calculate the density matrix $\rho$ of the two-qubit, bus-resonator system following the measurement pre-rotations.
Next, we use the calibrated readout errors for $Q$ and $A$ to calculate conditioned and unconditioned averages. For example,

\begin{widetext}
\[
\begin{array}{lcl}
\avg{\Mq}_{|\Ma=+1} &=&\frac{\FAz \left[  (2\FQz-1)\PAzQz -(2\FQo-1) \PAzQo\right] + \left(1-\FAo\right) \left[(2\FQz-1)\PAoQz -(2\FQo-1)\PAoQo\right]  }{\PAz \FAz + \PAo \left(1-\FAo\right)}, \\
\avg{\Mq}_{|\Ma=-1} &=&\frac{\FAo \left[  (2\FQz-1)\PAoQz -(2\FQo-1) \PAoQo\right] + \left(1-\FAz\right) \left[(2\FQz-1)\PAzQz -(2\FQo-1)\PAzQo\right]  }{\PAo \FAo + \PAz \left(1-\FAz\right)},\\
\avg{\Mq}&=&(2\FQz-1)\PQz -(2\FQo-1) \PQo,\\
\end{array}
\]
\end{widetext}
where
\[
\begin{array}{lcl}
p_{Ai}&\equiv&\tr\left[ \keta{i}\braa{i}\rho \right],\\
p_{Qj}&\equiv&\tr\left[ \ketq{j}\braq{j}\rho \right],\\
p_{Ai,Qj}&\equiv&\tr\left[ \ket{j_Q i_A}\bra{j_Q i_A}\rho \right]\\
\end{array}
\]
are the probabilities of $A$ being in $\keta{i}$,
$Q$ being in $\ketq{j}$, and $A$ and $Q$ being in $\ket{j_Q i_A}$, respectively, and $F_{Ai}$ and $F_{Qj}$ are the single-shot readout fidelities for $A$ and $Q$ calibrated in Fig.~\ref{Fig:Readout}.
Conditioned and unconditioned averages for $\Ma$ are similarly calculated.

\subsection{Quantum efficiency}
We consider the evolution of the qubit density matrix $\rhoQ$ conditioned on specific ancilla-measurement results:
\[
\rhoQ=
\left(
\begin{array}{cc}
\rho_{00} & \rho_{01}\\
\rho_{10} & \rho_{11}\\
\end{array}
\right)
\stackrel{\Ma=i}{\longrightarrow}
\rhoQ'=\left(
\begin{array}{cc}
\rho_{00}' & \rho_{01}'\\
\rho_{10}' & \rho_{11}'\\
\end{array}
\right).
\]
We define the quantum efficiency $\eta_i$ for measurement outcome $\Ma=i$ by
\[
\frac{|\rho_{01}'|}{\sqrt{\rho_{00}'\rho_{11}'}}=\eta_{i} \frac{|\rho_{01}|}{\sqrt{\rho_{00}\rho_{11}}},
\]
capturing the loss of information about the qubit as a result of the partial measurement~\cite{Korotkov08}.
For an ideal indirect measurement with perfect interaction and readout steps, it is straightforward
to show $\eta_{+1}=\eta_{-1}=1$. Including infidelity in the ancilla readout gives, for $\Ma=+1$,
\[
\rhoQ' \propto \FAz \Omega_{+1} \rhoQ \dagg{\Omega_{+1}} + \FAob \Omega_{-1} \rhoQ \dagg{\Omega_{-1}},
\]
where $\FAob\equiv 1-\FAo$.  The expression for $\Ma=-1$ is obtained by substituting $\FAz \rightarrow \FAzb$ and $\FAob \rightarrow \FAo$. The outcome-specific quantum efficiencies for our choice of qubit-ancilla interaction and measurement in the $\Xa$ basis become
\begin{widetext}
\[
\eta_{+1}=\frac{(\FAz+\FAob)\cos(\epsilon/2)}{\sqrt{\left[\FAz+\FAob + (\FAz-\FAob)\sin(\epsilon/2) \right]
\left[\FAz+\FAob - (\FAz-\FAob)\sin(\epsilon/2) \right]}}
\]
and
\[
\eta_{-1}=\frac{(\FAzb+\FAo)\cos(\epsilon/2)}{\sqrt{\left[\FAzb+\FAo -(\FAzb-\FAo)\sin(\epsilon/2)\right]\left[\FAzb+\FAo +(\FAzb-\FAo)\sin(\epsilon/2)\right]}}.
\]
These quantum efficiencies do not depend on the initial state of the qubit. The asymmetry in the ancilla readout fidelities causes $\eta_{+1}$ and $\eta_{-1}$ to differ. Including the residual excitation of the ancilla, $\PAe$, further reduces the quantum efficiencies to
\[
\eta_{+1}= \frac{(\FAz+\FAob)\cos(\epsilon/2)}
{\sqrt{\left[\FAz+\FAob + (1-2\PAe)(\FAz -\FAob)\sin(\epsilon/2)\right]\left[\FAz+\FAob - (1-2\PAe)(\FAz -\FAob)\sin(\epsilon/2)\right]}}
\]
and
\[
\eta_{-1}= \frac{(\FAzb+\FAo)\cos(\epsilon/2)}
{\sqrt{\left[\FAzb+\FAo - (1-2\PAe)(\FAzb -\FAo)\sin(\epsilon/2)\right]\left[\FAzb+\FAo + (1-2\PAe)(\FAzb -\FAo)\sin(\epsilon/2)\right]}}.
\]
\end{widetext}

Including decoherence of the ancilla, qubit and bus makes $\eta_{+1}$ and $\eta_{-1}$ dependent on the qubit input state. In this case we can average $\eta_i$ over the surface of the qubit Bloch sphere to arrive at single numbers. Lacking closed-form formulas, we rely on the master equation simulation to calculate
\[
\bar{\eta}_i\equiv\frac{\int \eta_i \sin(\theta) d\theta d\phi}{4\pi}.
\]

